\documentclass[12pt]{article}
\usepackage{float}
\usepackage[utf8]{inputenc}
\usepackage[margin=1in]{geometry}
\usepackage{amsmath}
\usepackage{amssymb}
\usepackage{color}
\usepackage{rotating}
\usepackage{booktabs} 
\usepackage{threeparttable}
\usepackage{array}    
\usepackage{natbib}   

\usepackage{graphicx}
\graphicspath{{images/}}
\usepackage{hyperref}
\title{A comparative study of two-sample hypothesis tests in the presence of long-term survivors}

\author{
Yu Bi$^{*}$, Durbadal Ghosh$^{*}$, Subodh Selukar \\[6pt]
\small Department of Biostatistics, St.\ Jude Children's Research Hospital, Memphis, TN, USA \\[6pt]
\small $^{*}$Equal contribution
}
\date{}

\begin{document}
\maketitle

\newpage
\begin{abstract}
Time-to-event data with long-term survivors (L-TS), subjects who never experience the event, have been reported in multiple areas of oncology as therapies have improved. Conventional two-sample tests ignore L-TS, but alternatives have been developed in the cure models literature. Because L-TS can induce non-proportional hazards (non-PH), non-PH candidates also exist. However, there has not been a comprehensive comparison of these candidates. Additionally, follow-up is an important consideration for data with L-TS, but there has been limited study of the impact of follow-up time on performance of two-sample tests with L-TS. We conducted a neutral simulation study of the impact of sample size and follow-up time on type I error and power across varying effect sizes for conventional methods, methods adapted for non-PH, and a correctly-specified parametric model. When one or both groups lack L-TS, log-rank tests and one non-PH method typically have the highest power, but order varies. Surprisingly, when both groups have L-TS, these tests have non-monotonic power as a function of follow-up time, while parametric models have monotonic increasing power and the highest power at the longest follow-up time. While absolute power differs, patterns over follow-up are consistent across sample sizes. To address this for practitioners, we devise a numerical approach to predict the potential for non-monotonicity during study planning. We conclude that naïve use of conventional methods can have counterintuitive properties in settings with L-TS, and this work provides knowledge and a tool to anticipate and address these issues.
\end{abstract}

\newpage
\section{Introduction}

Advances in cancer therapeutics have led to a growing number of clinical settings in which a substantial proportion of patients do not experience the event of interest in a practical timeline and are commonly referred to as long-term survivors (L-TS). For example, checkpoint inhibitors in advanced melanoma now produce 5-year overall survival rates approaching 50\% with combination therapy and 44\% with monotherapy \citep{Larkin2019, Wolchok2022}. More broadly, authors have used the Surveillance, Epidemiology, and End Results (SEER) database to document long-term survivorship across 42 cancer sites\citep{Tai2005}. Data with L-TS violate the fundamental assumption of conventional survival analysis that all subjects will eventually experience the event, suggesting that conventional methods may be limited in this setting.

Motivated by pediatric oncology, \citet{Sposto2002} suggested that cure models provide a coherent way to distinguish covariate effects on the proportion cured (L-TS who never experience the event of interest) from those on time to failure among non-L-TS. Two primary statistical frameworks have emerged to address cure modeling. Mixture cure models (MCM) explicitly separate cured and uncured subpopulations through $S(t) = \pi+ (1-\pi)S_u(t)$, where $\pi$ denotes cure probability and $S_u(t)$ represents survival for uncured patients \citep{Boag1949, Berkson1952, Kuk1992}. These have been substantially developed, including e.g., semiparametric approaches that allow flexible baseline hazards through nonparametric estimation \citep{Sy2000, Peng2000} and Bayesian implementations by \citet{Chen1999}. Non-mixture models take a different approach by bounding the cumulative hazard via $S(t) = \exp(-\eta F(t))$, where $F(\cdot)$ is a distribution and these yields a cure fraction $\exp(-\eta)$ \citep{Tsodikov1998, Tsodikov2003}. Recent theoretical advances have provided more generalized frameworks for cure models \citep{Patilea2020}.

Non-proportional hazards (NPH) emerge naturally in data with L-TS. For example, based on the mixture framework, when two treatment groups have identical uncured survival but different cure fractions, survival curves diverge over time. When groups have different uncured survival but similar cure fractions, survival curves converge to the same plateau.
NPH patterns have been observed in real-world immunotherapy trials, for example, where delayed treatment effects occur with minimal early separation preceding substantial divergence after immune activation \citep{Wei2020, Wu2022}. These NPH patterns invalidate the proportional hazards assumptions that underlie Cox regression and standard log-rank tests. \citet{Kim2012} introduced a three-component cure model for NPH designs and also showed that misspecifying the alternative can underpower trials. Alternative approaches such as restricted mean survival time (RMST) and survival probability differences have gained attention for handling NPH without relying on hazard-based assumptions \citep{Royston2013, Uno2014}, but they require setting appropriate time horizons. Weighted log-rank tests can address specific NPH patterns using Fleming-Harrington weights $w(t) = \hat{S}(t)^{\rho}[1-\hat{S}(t)]^{\vartheta}$, where $\rho$ and $\vartheta$ control emphasis on early versus late differences \citep{Fleming1991}. MaxCombo methods combine multiple weighted log-rank tests to maintain power across different NPH scenarios \citep{Royston2016, Mukhopadhyay2022}. Recently, \citet{Klinglmueller2025} conducted a comprehensive simulation study of candidate methods under varying non-PH settings. They found advantages to different methods based on power (log-rank-type tests and model-based methods) versus interpretability (RMST). Notably, they did not include data-generating mechanisms with L-TS. Building on their study, this investigation focuses on evaluating the type-1 error and power of hazard-based methods under various settings with L-TS.

Conventional studies typically assess data maturity through censoring distributions and median follow-up \citep{Betensky2015, Schemper1996, Clark2003, Gebski2018}. Follow-up in data with L-TS requires special consideration because of the need for sufficient follow-up to identify cure models, and insufficient follow-up can induce unpredictable bias in cure model estimates \citep{Othus2020}. \citet{Maller1992}, \citet{Maller1996} and \citet{Selukar2023} developed procedures for the adequacy of cure models for a given dataset. Across populations, \citet{Tai2005} found that minimum requirements ranged from 2.6 years for pancreatic cancer to effectively unmeasurable durations exceeding 25 years for indolent cancers such as thyroid and prostate cancer.

The performance of different statistical methods across varying follow-up durations in data with L-TS remains largely unexamined, as cure methodology frequently assumes sufficient follow-up. In general, previous investigations have addressed either NPH scenarios \citep{Kim2012, Xiong2017, Wu2022} or follow-up requirements \citep{Tai2005, Selukar2023} separately. This gap is problematic because many oncology trials may conclude before follow-up is sufficient for cure modelling, while simultaneously encountering NPH from delayed treatment effects or differential cure fractions. Understanding how limited follow-up affects available methods is critical for appropriate trial design and analysis.

While \citet{Peng2021} examined cure models under specific scenarios with delayed treatment effects through simulations, this investigation provides the first comprehensive study comparing methods potentially relevant for conducting two-sample hypothesis testing in data with L-TS, while explicitly examining how changing follow-up duration affects their performance. Our dual focus addresses an important gap in the literature. We systematically evaluate how method performance in terms of power and type I error changes as the follow-up extends from early time points through longer follow-up. Given the widespread use of the conventional log-rank test, we also develop a tool to predict the shape of its power over follow-up time in the presence of L-TS.
Section 2 describes the candidate statistical methods, the simulation design using mixture cure models, and the theoretical derivation for predicting the shape of the power of the log-rank test statistic. Section 3 presents the simulation results for type I error and power for  candidate methods and evaluates the relationship between the proposed approach for predicted power and the empirical power. Section 4 contextualizes these results using reconstructed data from a neuroblastoma trial. Finally, Section 5 discusses the implications of these findings for trial design and the limitations of the study.

\section{Methods}

The primary goal of this manuscript is to assess the performance, measured by estimated type I error and statistical power, of two-sample hypothesis tests in data with L-TS. We evaluated the impact of sample size and follow-up time across a range of effect sizes on the performance of the two-sample hypothesis tests. We focused on two-sample comparisons with hypotheses
\begin{equation}
H_0: S_0(t) = S_1(t) \text{ for all } t \geq 0 \textrm{ vs. } H_a: S_0(t) \neq S_1(t) \text{ for some } t \geq 0
\end{equation}
where $S_0(t)$ is the survival function for the control group and $S_1(t)$ is the survival function for the treatment group.

\subsection{Candidate methods}
A variety of methods have been developed for two-sample survival comparisons under different data structures (Table \ref{tab:survival_comparison}). We aimed to conduct a neutral simulation study of the available candidates.

The conventional nonparametric approach is the Log-Rank (LR) test, which is the most powerful rank-based test under the proportional hazards (PH) assumption. The LR test assigns equal weight ($w=1$) to all events across the follow-up period and assumes that all subjects will eventually experience the event (i.e., no L-TS exist).

\citet{Boag1949} proposed the mixture cure model, which assumes two distinct subpopulations: a cured subpopulation of proportion $\pi$ who will never experience the event, and an uncured subpopulation of proportion ($1 - \pi$) whose survival follows a proper distribution function denoted by $S_u(t)$. Mixture cure models have been substantially developed in the broader cure models literature \citep{Amico2018, Peng2021}. We use the correctly-specified parametric mixture cure model as a benchmark method optimized for our data-generating mechanism.

Weighted Log-Rank (WLR) tests apply different weights to the log-rank statistic to improve detection of hazard differences under NPH or at specific time periods. Using the Fleming-Harrington $(G^{\rho, \vartheta})$ weight family,
\begin{equation}
w = \hat{S}(t)^\rho (1-\hat{S}(t))^\vartheta
\end{equation}

with the Kaplan-Meier estimate of the pooled survival function $\hat{S}(t)$ and non-negative parameters $\rho \geq0$ and $\vartheta \geq0$ controlling the placement of weight, we include three variants: (1) Early WLR ($G^{1,0}$), which emphasizes early hazard differences; (2) Late WLR ($G^{0,1}$), which emphasizes late hazard differences; and (3) Optimal LR, where weights are proportional to the magnitude of the log-hazard ratio, shown to be the most powerful weighted log-rank test \citep{Schoenfeld1981}.

Because WLR tests require pre-specification of weight parameters, adaptive weighted log-rank tests have been developed. We consider the Yang-Prentice (YP) method proposed by Yang and Prentice \citep{YangPrentice2010}. This method models a flexible hazard function
\begin{equation}
\lambda(t \mid x=1) = \frac{\theta_E \theta_L}{\theta_E + (\theta_L - \theta_E) S(t \mid x=0)} \; \lambda(t \mid x=0),
\end{equation}
where $S(t\mid x=0)$ is baseline survival function in the control group, $\theta_E$ and $\theta_L$ represent the short-term and long-term hazard ratios, respectively. When $\theta_E=\theta_L$, the hazard model reduces to the proportional hazards model. By allowing different $\theta_E$ and $\theta_L$, the model can capture various NPH patterns, including crossing survival curves, disappearing treatment effects, or delayed treatment effects. We also evaluated the adaptive Two-Stage (TS) method \citep{QiuSheng2008}, which applies a standard log-rank test in the first stage and then selects Fleming-Harrington weights via bootstrap in the second stage. Both approaches adapt to the underlying data structure without requiring pre-specification of the timing of treatment effects.

The mathematical and procedural details of these tests are listed in the Appendix D.

\subsection{Simulation design}
Survival data were generated randomly from a mixture cure model, including both cured and uncured subpopulations $S(t) = \pi + (1 - \pi) S_u(t)$, where $\pi$ is the incidence (cure or L-TS proportion) and $S_u(t)$ is the latency (survival function of uncured or non-L-TS subpopulation).

The control group served as the reference arm, defined by a cure fraction $\pi_0$ and an uncured survival function $S_{u,0}$ following a pre-specified distribution. Treatment differences were introduced via two separate effect sizes, odds ratio for the cure fraction ($OR$) and hazard ratio among the uncured subpopulation ($HR_u$). The $OR$ quantifies the treatment effect on the odds of being cured with formula $\operatorname{logit}(\pi_1) - \operatorname{logit}(\pi_0) = \ln(\mathrm{OR})$. 

The hazard function for the uncured subpopulation, $h_u(t)$, is related to $S_u(t)$ by $h_u(t) = -\frac{d}{dt}\log S_u(t).$ Under a PH assumption for the uncured subpopulation, we have $h_{u1}(t)=HR_u\cdot h_{u0}(t)$, which implies that the survival functions for the uncured fraction are related to $S_{u,1}(t)=[S_{u,0}(t)]^{HR_u}$. For log-logistic and Gamma distributions, the accelerated failure time (AFT) model is applied and the $HR_u$ values were mapped to corresponding time ratio for uncured subpopulation ($TR_u$) to maintain comparable effect sizes across settings. For reference, we visualize several data-generating mechanisms used in the main text results in the Supplement.

In the simulation, $\tau$ represents the end of the follow-up. We varied $\tau$ based on selected quantiles of the uncured survival distribution in the control group, which reflects decreasing proportion of the uncured subpopulation of the control group remaining at risk. We rounded the $\tau$ to the nearest quarter-year to approximate standard calendar time practices in clinical studies. The simulation was conducted among the systematic variation of the factors in Table \ref{tab:sim_parameters}. For each combination of the factors, we generated 10,000 simulated datasets. Below is the survival time and censoring process:

\begin{enumerate}
    \item True survival time ($T$): We simulated event times separately by arm according to the specified cure fraction and uncured survival function with the inverse CDF method.
    \item Censoring mechanism: Consistent with clinical trial recruitment, subjects were accrued over a fixed calendar accrual time ($\tau_{acc}$), defined as half of the 75th percentile of the uncured survival distribution in the control group. Type I right-censoring was applied in the simulation. The censoring time $C$ for each subject was independently generated from a uniform distribution ranging from $\tau_{acc}$ to the specified study termination time $\tau$. 
    \item Observed data: The data available for hypothesis testing in each simulated dataset was $T_{obs}=\min(T, C)$ and event indicator $\delta = I(T \leq C)$ and the group indicator.
\end{enumerate}

 The performance of the methods was evaluated using type I error and statistical power. We computed the Monte Carlo estimate of the hypothesis test rejection rate for each method as the proportion of simulated datasets for which the two-sided p-value is less than 0.05. This rejection rate estimates the type I error where both $OR=1$ and $HR_u=1$. The rejection rate estimates the power where $OR>1$ or $HR_u<1$ or both. In this study, we did not evaluate settings where the $OR$ and $HR_u$ suggested different directions of effect (i.e., in all settings the survival function of treatment group was equal to or uniformly higher than the survival function of control group). All data generation and subsequent analysis were performed on a High Performance Computing (HPC) cluster and recorded with random number generation seed and index for each simulation scenario to ensure the reproducibility and trackability of the simulation results.

See packages and code details in Appendix E. During execution on the HPC, a small number of simulations resulted in numerical failures. All failed cases were systematically documented, recording the simulation index, data, test applied, and error details. After investigation, these failures were attributed to zero events in one or more groups. Across all settings, rates of failure were $\leq 0.2\%$. Performance metrics were summarized across replicates without numerical failure.

\subsection{Numerical prediction of power curve of log-rank test with L-TS}
To investigate the expected properties of log-rank test statistics, we derived the expected value of the numerator of the log-rank test statistic at fixed follow-up times. We use this as a numerical tool to predict the shape of the power curve across follow-up time for the log-rank test in the presence of L-TS.

Let $t$ denote time and let $\tau$ denote the follow-up time at which the test is evaluated. For groups $g\in\{0,1\}$, define the counting process $n_g(t)$ (number of observed events in group $g$ by time $t$), the at-risk process $Y_g(t)$ (number still under observation at time $t$), and the pooled quantities $n(t)=n_0(t)+n_1(t)$ and $Y(t)=Y_0(t)+Y_1(t)$. The log-rank score accumulated up to follow-up $\tau$ is
\begin{equation}
U(\tau)=\int_0^\tau\Big\{dn_1(t)-\frac{Y_1(t)}{Y(t)}\,dn(t)\Big\}.
\label{eq:U}
\end{equation}
Under independent censoring, the expected log-rank score is approximately a weighted integral of the hazard difference between arms:
\begin{equation}
\mathbb{E}\{U(\tau)\}\approx \int_0^\tau w(t)\{h_1(t)-h_0(t)\}\,dt,
\label{eq:LRmean}
\end{equation}
where $h_g(t)$ is the hazard function for group $g$. Under balanced allocation, large samples, and uniform censoring $C\sim\mathrm{Uniform}(\tau_{\mathrm{start}},\,\tau_{\mathrm{end}})$ with survival function $G(t)=(\tau_{\mathrm{end}}-t)/(\tau_{\mathrm{end}}-\tau_{\mathrm{start}})$ for $\tau_{\mathrm{start}}\le t\le\tau_{\mathrm{end}}$, this weight $w(t)$ simplifies to:
\begin{equation}
w(t)\;=\;G(t)\,\frac{2S_0(t)S_1(t)}{S_0(t)+S_1(t)} ,
\label{eq:wdef}
\end{equation}
where $S_g(t)$ is the survival function for group $g$. Further details are provided in the Supplement. In general, the distribution of $G(t)$ can be chosen to be any distribution, whether informed by prior information or other considerations; the shape of $A(\tau)$ (as defined below) will change accordingly, but $A(\tau)$ can still be used as a numerical tool for informing the shape of the power of the log-rank test. Further details are provided in the Supplement.

Under the mixture cure model, the survival and hazard functions for group $g$ are
\begin{equation}
S_g(t)=\pi_g+(1-\pi_g)S_{u,g}(t),
\label{eq:mixtureS}
\end{equation}
\begin{equation}
h_g(t)=\frac{(1-\pi_g)f_{u,g}(t)}{S_g(t)} ,
\label{eq:mixtureh}
\end{equation}
where $\pi_g$ is the cure fraction, $S_{u,g}(t)$ is the survival function of uncured subjects, and $f_{u,g}(t)=-S_{u,g}'(t)$ is their density. As $t$ increases, $S_g(t)\to\pi_g$ and $h_g(t)\to 0$. We define the cumulative hazard contrast
\begin{equation}
\Delta H(t)=H_1(t)-H_0(t),
\qquad
H_g(t)=-\log S_g(t).
\label{eq:DeltaH}
\end{equation}
We then define the \emph{weighted average hazard difference}
\begin{equation}
A(\tau)\;=\;\int_0^\tau w(t)\,\{h_1(t)-h_0(t)\}\,dt ,
\label{eq:Adef}
\end{equation}
so that $\mathbb{E}\{U(\tau)\}\approx A(\tau)$. The standardized log-rank test statistic is $Z(\tau) = U(\tau)/\sqrt{V(\tau)}$, where $V(\tau) = \mathrm{Var}\{U(\tau)\}$ is the variance of the log-rank score under the null hypothesis. Because the power of a two-sided test depends on $|Z(\tau)|$, qualitative changes in power as a function of $\tau$ are driven by changes in the numerator $A(\tau)$ provided that $V(\tau)$ changes slowly relative to $A(\tau)$. Under this assumption, the sign and qualitative pattern of the standardized log-rank statistic track those of $A(\tau)$ as follow-up increases. We evaluate $A(\tau)$ and $\Delta H(t)$ numerically on a dense grid of follow-up quantiles (more details are given in the Supplement).

\section{Results}
\subsection{Simulation results}
We conducted a comprehensive set of simulations using Weibull, gamma and log-logistic uncured distributions. The main text reports results for the Weibull distribution and other results are provided in the supplementary material because of similar patterns using gamma and log-logistic distributions as the Weibull distribution. Below, we summarized the main findings and discussed how method performance varies with key factors. For illustration purposes, the main text focuses on the LR test, mixture cure model, optimal LR test and adaptive YP test. These selected methods perform similarly to or better than other tests within the same modeling framework. The adaptive YP test performs comparably or better to the adaptive TS test across all settings, and the optimal LR test frequently outperforms early- and late- WLR tests. Figures with all results are shown in the supplementary material.

Type I error was generally well-controlled across all methods and settings (Supp Figure~7). However, the optimal log-rank test exhibited an elevated Type I error (0.051-0.108) when no cure fraction was present in both groups and follow-up longer than 0.95 quantile follow-up time, regardless of sample size. Because it maintained Type I error for non-zero cure fractions and otherwise frequently outperformed other weighted log-rank tests, we included it as a representative candidate for weighted log-rank tests in the rest of the main text results. Similarly, the adaptive YP test can have modestly inflated Type I error (up to 0.082), but it outperformed adaptive TS, so we include it as a representative candidate of adaptive tests throughout the remainder of the main text results.

\begin{figure}[h]
    \centering
    \includegraphics[width=1\textwidth]{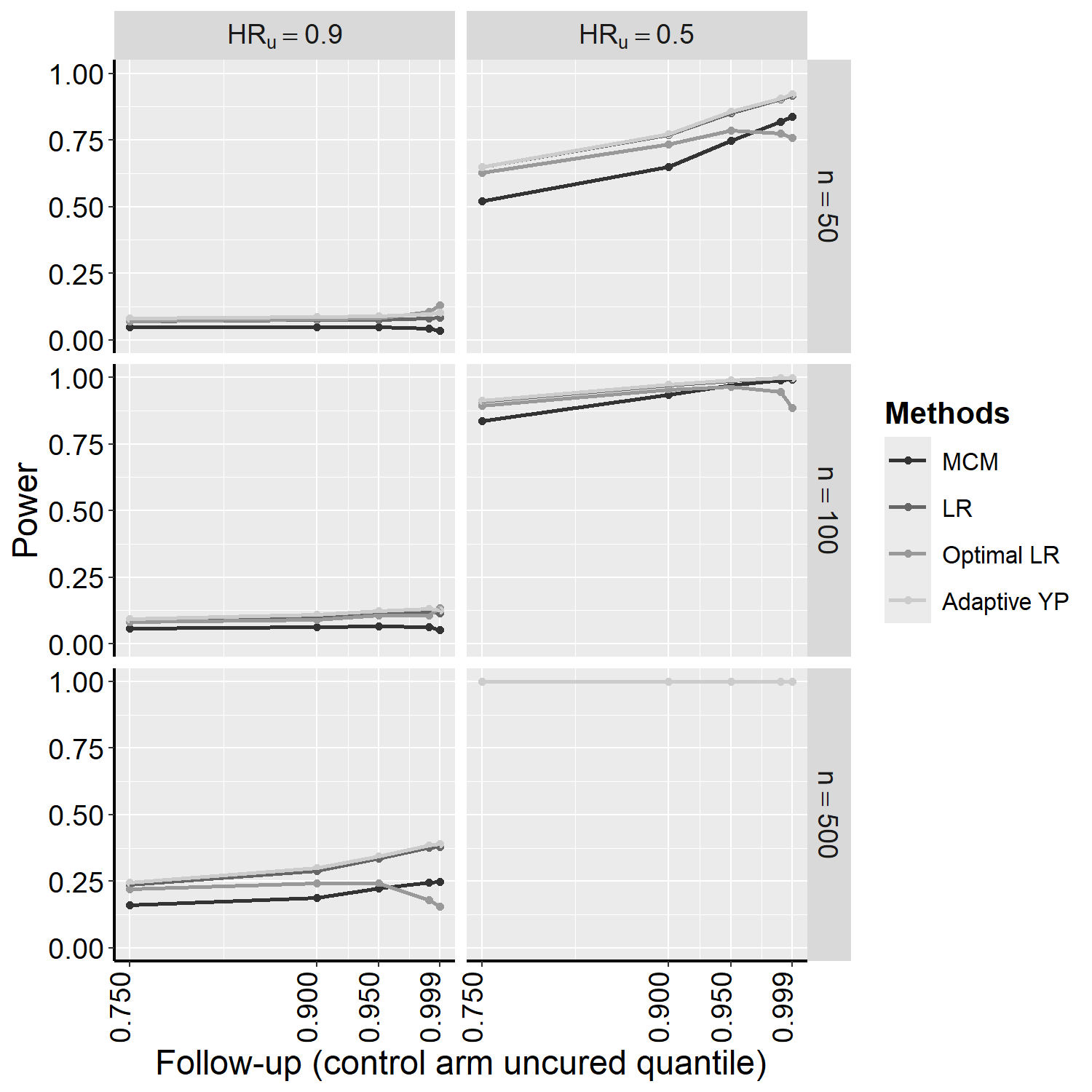}
   \caption{Rejection rate (power) of four two-sample survival tests as a function of follow-up percentile under Weibull uncured survival with no cured fraction in either group ($\pi_0 = \pi_1 = 0$). Columns correspond to the hazard ratio for the uncured distribution ($HR_u = 0.9$ and $HR_u = 0.5$). Rows correspond to sample sizes $n = 50$, $100$, and $500$ per group. The four methods shown are the log-rank test (LR), mixture cure model (MCM), optimal log-rank test (Optimal LR), and adaptive Yang-Prentice test (Adaptive YP). Complete results for all methods are shown in the supplementary material.}
   \label{fig:power_nocure}
\end{figure}

\begin{figure}[h]
    \centering
    \includegraphics[width=1\textwidth]{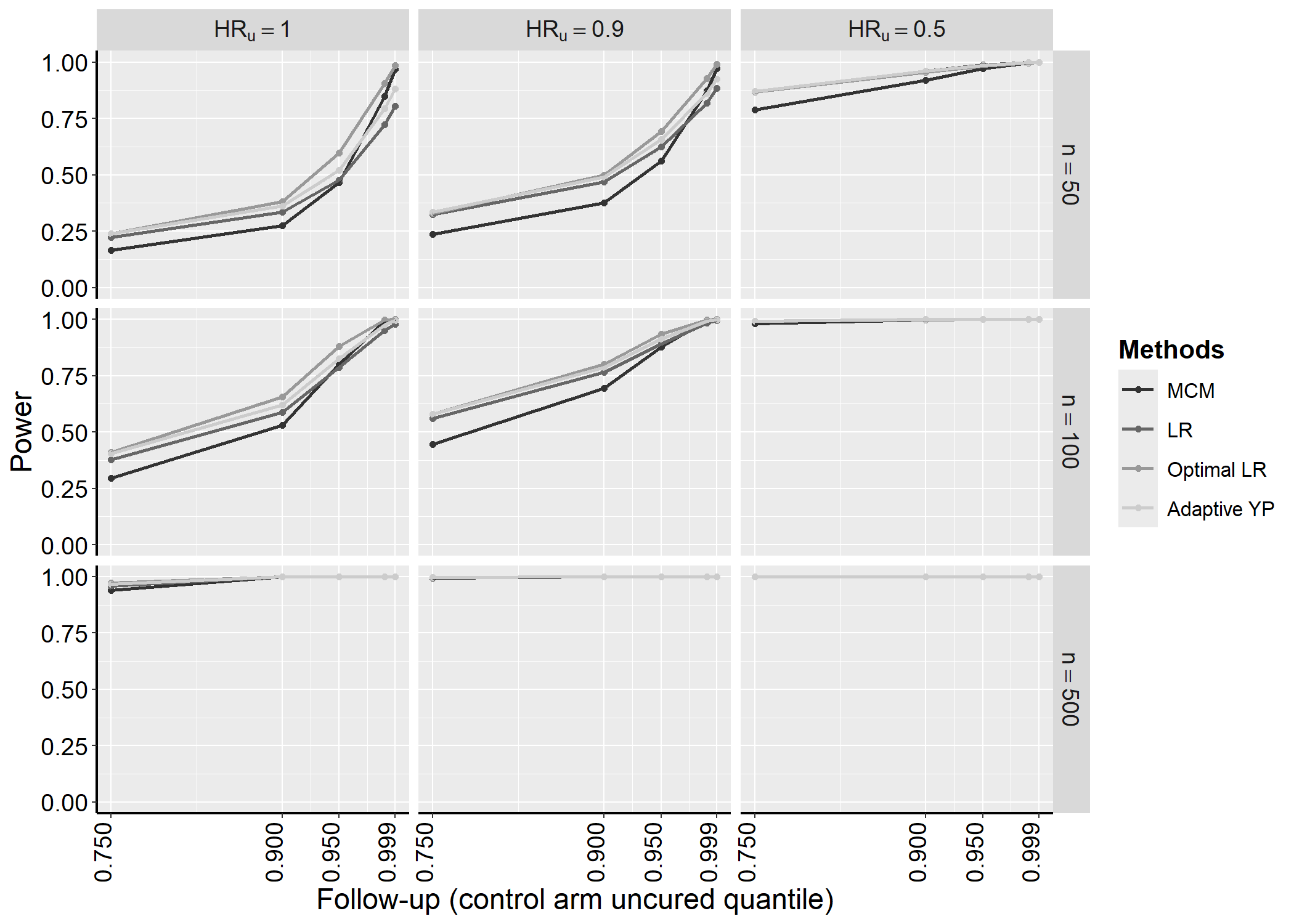}
    \caption{Rejection rate (power) of four two-sample survival tests as a function of follow-up percentile under Weibull uncured survival with no cured fraction in the control group ($\pi_0 = 0$) and a fixed cured fraction in the treatment group ($\pi_1 = 0.2$). Columns correspond to the hazard ratio for the uncured distribution ($HR_u = 1$, $0.9$, and $0.5$). Rows correspond to sample sizes $n = 50$, $100$, and $500$ per group. The four methods shown are the log-rank test (LR), mixture cure model (MCM), optimal log-rank test (Optimal LR), and adaptive Yang-Prentice test (Adaptive YP). Complete results for all methods are shown in the supplementary material.}

    \label{fig:power_1cure}
\end{figure}

In scenarios without L-TS in one or both groups (Figures \ref{fig:power_nocure} and \ref{fig:power_1cure}), the log-rank test, optimal LR test, and adaptive YP test were typically the most powerful, though their relative performance varied by setting. When both groups had no cured subjects, and we varied only the $HR_u$ (a PH scenario, Figure~\ref{fig:power_nocure}), the log-rank and adaptive YP tests generally showed the highest power, which increased with longer follow-up and sample size, as expected. The power of the optimal LR test did not increase monotonically with longer follow-up and decreased after following 95th percentile of control group uncured distribution. The mixture cure model had limited power with $HR_u=0.9$ across sample sizes, but power was comparable to the log-rank and adaptive YP tests with $HR_u=0.5$ and long follow-up exceeding 90th percentile and $n\geq 100$ per group. Alternatively, when the treatment group had a low cure proportion ($\pi_1=0.2$, Figure \ref{fig:power_1cure}), all methods performed well, with power increasing as effect size, sample size, and follow-up time increased. The mixture cure model had the lowest power at the $75th$ and $90th$ percentiles of follow-up among these scenarios, but its performance approached that of the other tests with longer follow-up.

\begin{figure}[h]
    \centering
    \includegraphics[width=1\textwidth]{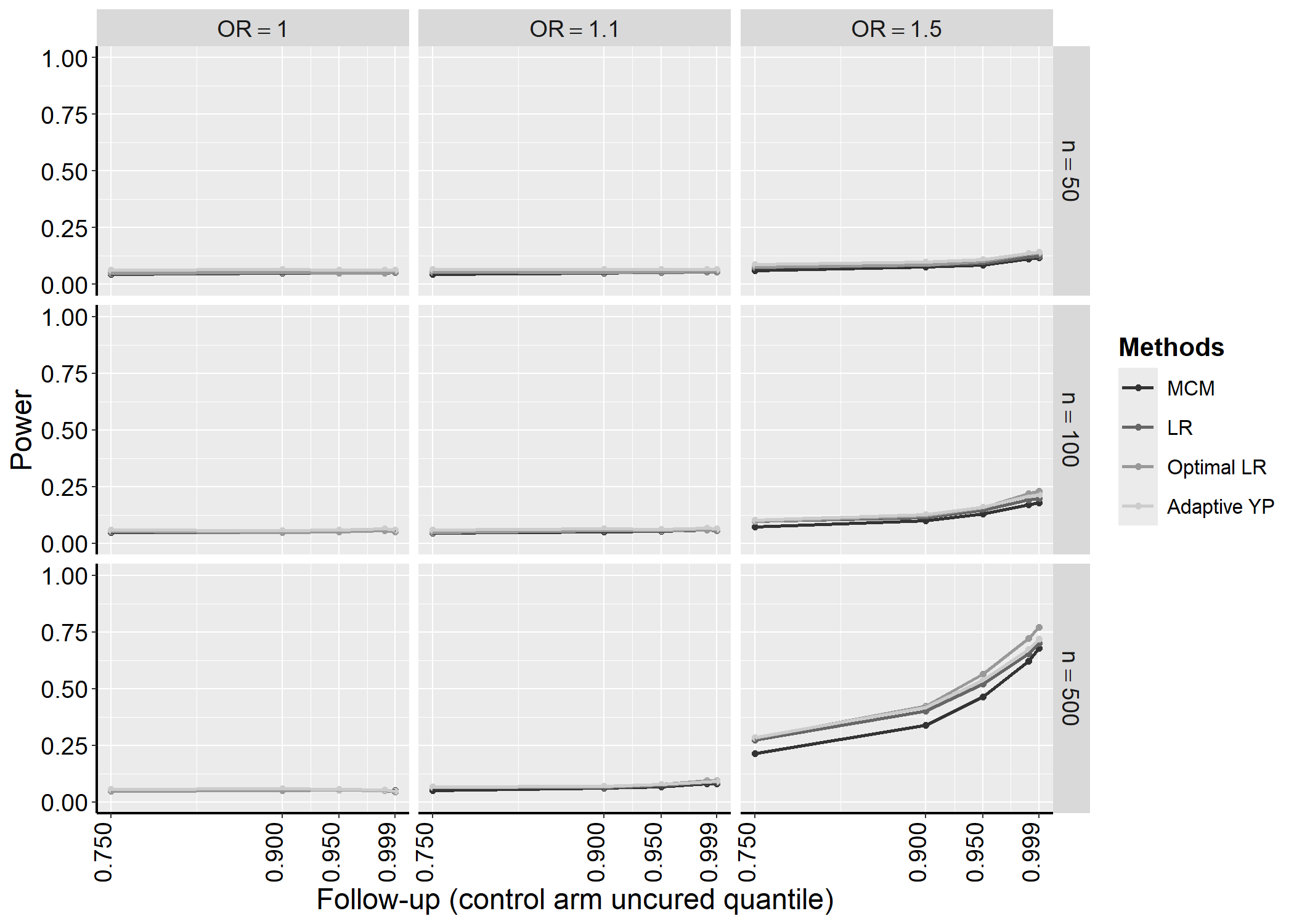}
    \caption{Rejection rate (power) of four two-sample survival tests as a function of follow-up percentile under Weibull uncured survival with cured fractions in both groups ($\pi_0 = 0.2$) and identical uncured survival distributions ($HR_u = 1$). Columns correspond to the odds ratio for the cured fraction ($OR = 1$, $1.1$, and $1.5$). Rows correspond to sample sizes $n = 50$, $100$, and $500$ per group. The four methods shown are the log-rank test (LR), mixture cure model (MCM), optimal log-rank test (Optimal LR), and adaptive Yang-Prentice test (Adaptive YP). Complete results for all methods are shown in the supplementary material.}

    \label{fig:power_2cure_iden}
\end{figure}

\begin{sidewaysfigure}[h]
    \centering
    \includegraphics[width=1\textwidth]{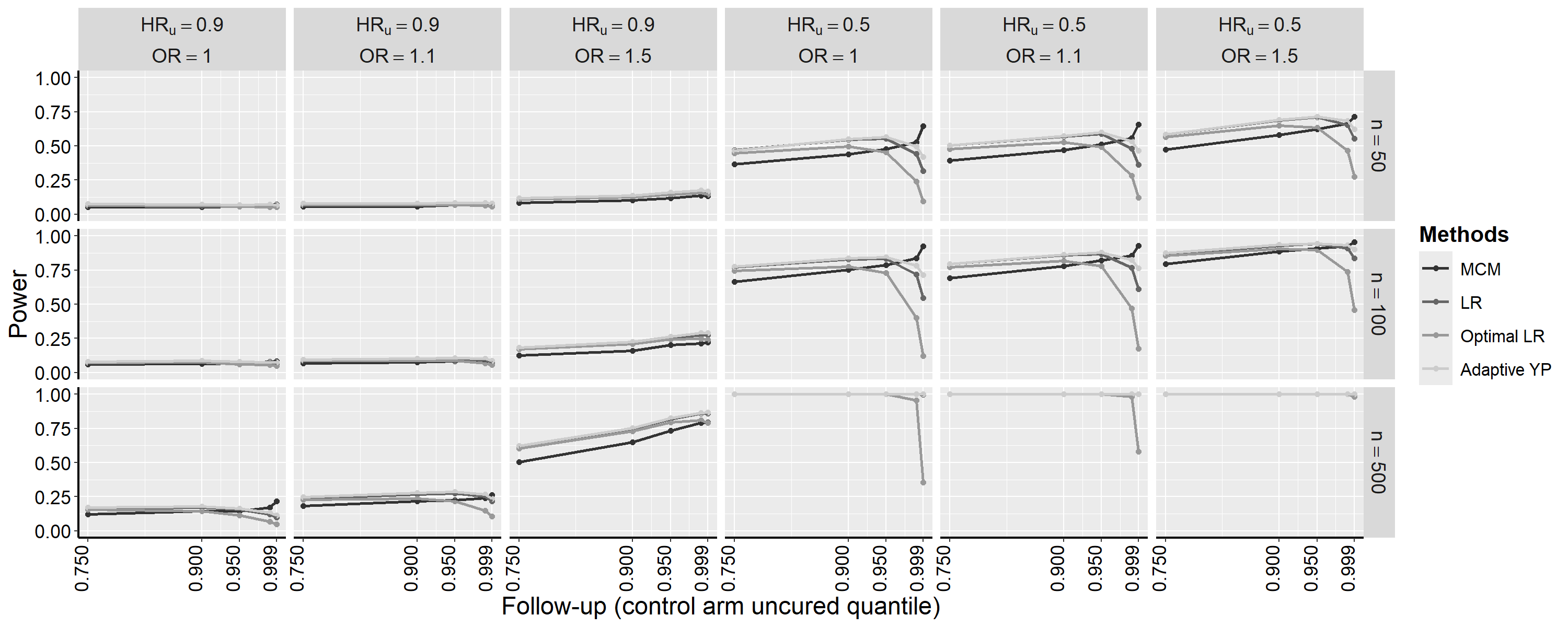}
    \caption{Rejection rate (power) of four two-sample survival tests as a function of follow-up percentile under Weibull uncured survival with cured fractions in both groups ($\pi_0 = 0.2$) and differing uncured survival distributions. Columns correspond to combinations of the hazard ratio for the uncured distribution ($HR_u = 0.9$ and $0.5$) and the odds ratio for the cured fraction ($OR = 1$, $1.1$, and $1.5$). Rows correspond to sample sizes $n = 50$, $100$, and $500$ per group. The four methods shown are the log-rank test (LR), mixture cure model (MCM), optimal log-rank test (Optimal LR), and adaptive Yang-Prentice test (Adaptive YP). Complete results for all methods are shown in the supplementary material.}

    \label{fig:power_2cure_2uncure}
\end{sidewaysfigure}

When both groups included L-TS (Figures \ref{fig:power_2cure_iden} and \ref{fig:power_2cure_2uncure}), patterns differed based on whether there was an effect on the uncured ($HR_u\neq 1$). When both groups had identical uncured survival distributions and varying effect on the cure fraction (Figure \ref{fig:power_2cure_iden}), all methods exhibited low power, even with large effect sizes ($OR=1.5$) and large sample sizes ($n=500$ per arm). In scenarios with varying effects on the uncured survival (Figure \ref{fig:power_2cure_2uncure}), the log-rank, the optimal weighted log-rank, and adaptive YP tests displayed non-monotonic power patterns, with power increasing initially but decreasing after the 90th percentile. On the other hand, the mixture cure model exhibited a monotonic increase in power, achieving the highest power with the longest follow-up. While the power values differed, patterns with increasing follow-up were consistent across different sample sizes and effects on the cure fraction.

\subsection{Comparing the predicted and empirical power over follow-up time for the log-rank test with L-TS}
\begin{figure}[!t]
  \centering
  \includegraphics[width=\textwidth]{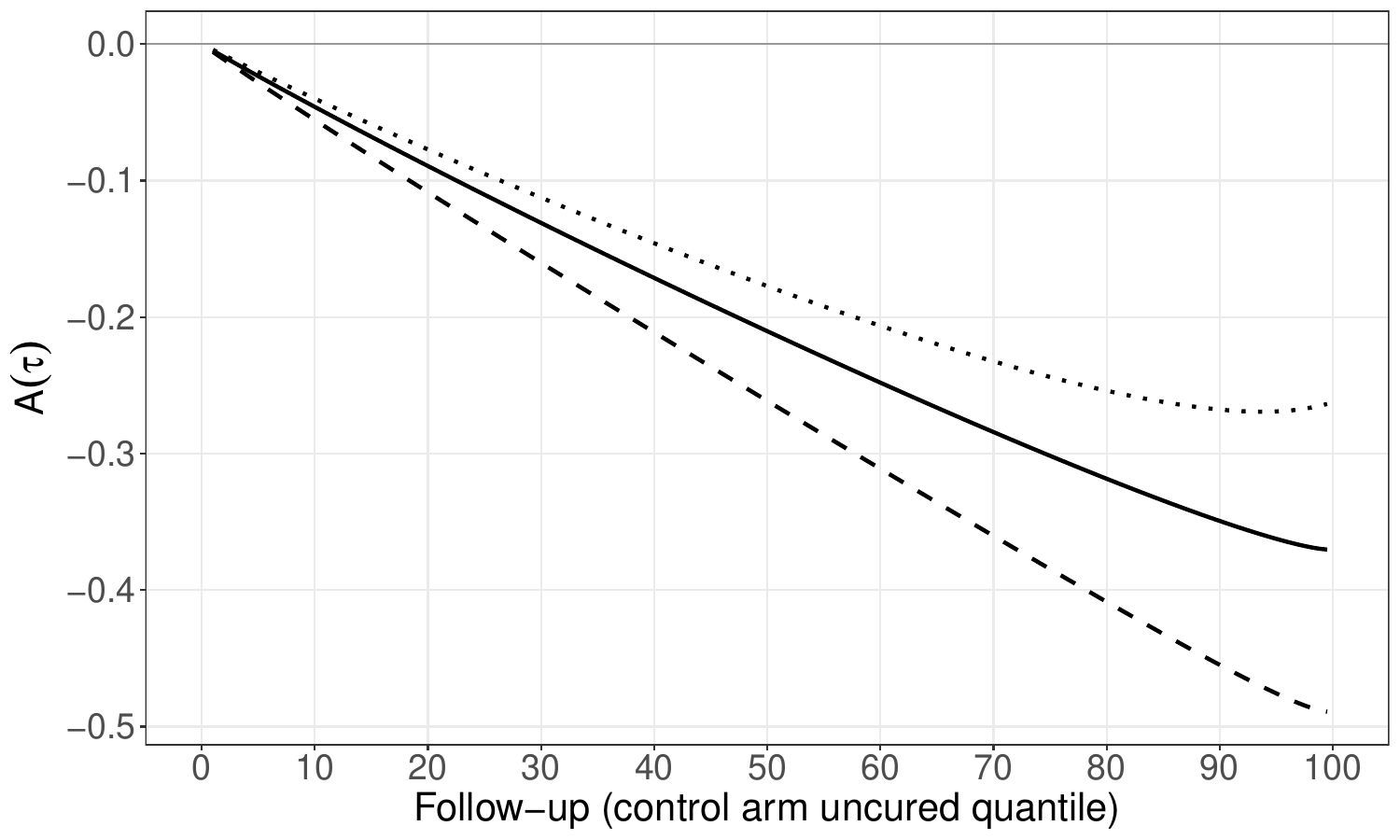}
\caption{Weighted average hazard difference \(A(\tau)\) under weight \(w(\tau)\) (defined in \eqref{eq:wdef}) plotted against follow up percentile \(\tau\) on the control uncured Weibull\((2,1)\) scale. The solid line corresponds to \(HR_u=0.5\) with \(\pi_0=0\) and \(\pi_1=0\). The dashed line corresponds to \(HR_u=0.5\) with \(\pi_0=0\) and \(\pi_1=0.2\). The dotted line corresponds to \(HR_u=0.5\) with \(\pi_0=0.2\) and \(OR=1.5\). The solid and dashed curves are monotone decreasing on this scale.}
  \label{fig:fig5}
\end{figure}

Figure~\ref{fig:fig5} displays the weighted average hazard difference \(A(\tau)\) defined in \eqref{eq:Adef} under the weight \(w(\tau)\) defined in \eqref{eq:wdef}. The horizontal axis is the follow-up percentile \(\tau\) on the control Weibull\((2,1)\) scale. The solid line corresponds to \(HR_u=0.5\) with \(\pi_0=0\) and \(\pi_1=0\). The dashed line corresponds to \(HR_u=0.5\) with \(\pi_0=0\) and \(\pi_1=0.2\). The dotted line corresponds to \(HR_u=0.5\) with \(\pi_0=0.2\) and \(OR=1.5\). By \eqref{eq:LRmean} together with \eqref{eq:Adef} the sign and the trajectory of \(A(\tau)\) track the log-rank signal at the same follow up, so we use the power plots for the same scenarios to check the empirical relationship between \(A(\tau)\) and power.

We see that this numerical approach appropriately predicts the empirical power, where empirical power is larger as $A(\tau)$ is further away from zero. In the no L-TS setting ($HR_u=0.5$, $\pi_0=\pi_1=0$) and the treatment-only L-TS setting ($HR_u=0.5$, $\pi_0=0$, $\pi_1=0.2$), the curve $A(\tau)$ decreases monotonically away from zero, mirroring the monotonic increase of the corresponding power curve over follow-up. In the both-groups L-TS setting with \(HR_u=0.5\), \(\pi_0=0.2\) and \(OR=1.5\) the dotted curve is non monotone. It starts at zero, decreases, reaches a minimum around the 95th percentile, and then increases while remaining below zero. The corresponding power curves are also non-monotone. They increase, attain a maximum, and then decrease as \(A(\tau)\) moves back toward zero.

Other scenarios appear in the Supplement. In every case, when \(A(\tau)\) is monotone, the power is monotone, and when \(A(\tau)\) is non-monotone, the power is non-monotone. The Supplementary Material also includes an example where \(A(\tau)\) stays negative and attains its minimum near the fiftieth percentile. In that example, the power rises early, peaks near the fiftieth percentile, and then declines. 

\section{Data Example}

\begin{figure}[h]
    \centering
    \includegraphics[width=0.95\textwidth]{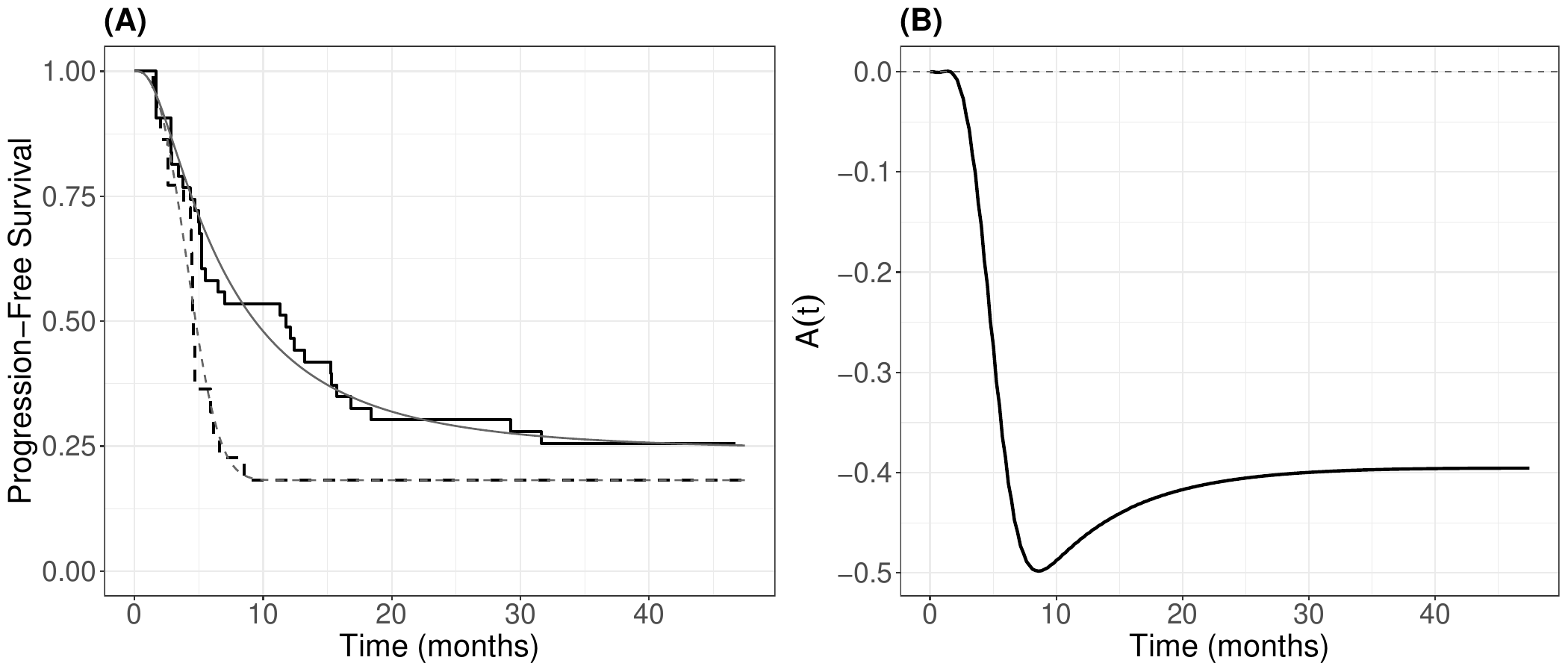}
     \caption{BEACON-Immuno trial analysis. Panel~A: Thick solid step curve is the Kaplan-Meier estimate for the dB plus chemotherapy arm; thick dashed step curve is the Kaplan-Meier estimate for the chemotherapy-alone arm. Thin smooth curves are estimated mixture cure model survival functions (solid: dB arm, $\hat{\pi}_1=0.24$; dashed: chemotherapy-alone arm, $\hat{\pi}_0=0.18$).  Panel~B: $A(t)$ computed from these fitted survival functions using the weight $w(t)$ defined in \eqref{eq:wdef}. The curve reaches its minimum near $t\approx 9$ months before increasing and plateauing near $A(t) \approx -0.39$ for $t > 30$ months, predicting non-monotonic log-rank power.}

    \label{fig:km_beacon}
\end{figure}

The BEACON-Immuno trial (NCT02308527) compared dinutuximab beta (dB) plus temozolomide-based chemotherapy versus chemotherapy alone in children with relapsed high-risk neuroblastoma \citep{Gray2026}. We reconstructed individual patient-level data from the published progression-free survival Kaplan-Meier curves using the WebDigitizer-based tool available at \url{https://biostatistics.mdanderson.org/shinyapps/IPDfromKM/}. The dataset includes 65 patients: 43 in the dB arm (32 events) and 22 in the chemotherapy-alone arm (18 events). Figure~\ref{fig:km_beacon}A displays the Kaplan-Meier curves, with both arms exhibiting plateaus suggesting L-TS.

We applied the RECeUS-AIC method \citep{Selukar2023} to assess cure model appropriateness. Per \citet{Kouadio2025}, we evaluated if at least one of the arms has evidence of being appropriately modeled with a cure model, and we see that the chemotherapy-alone arm satisfies the RECeUS criteria, supporting the use of cure modeling for these data. (Further details are in the supplement.)

To illustrate how our numerical approach can inform future study planning for these patients and this agent (e.g., confirmatory trial development), we apply results developed herein. We treat the fitted mixture cure models as the data-generating mechanism. For the chemotherapy-alone arm, we use a Weibull uncured distribution with shape $k = 2.64$ and scale $\lambda = 4.85$ (estimated from the data) and cure fraction $\pi_0 = 0.18$. For the dB arm, we use a log-normal uncured distribution with $\mu = 1.87$ and $\sigma = 0.88$ (estimated from the data), where $\mu$ and $\sigma$ are the mean and standard deviation of log-survival and cure fraction $\pi_1 = 0.24$. Figure~\ref{fig:km_beacon}B displays $A(t)$ defined in \eqref{eq:Adef}, computed from these fitted survival functions using the weight $w(t)$ defined in \eqref{eq:wdef} and the censoring distribution $G(t)$ with $\tau_{\mathrm{start}}=0$ and $\tau_{\mathrm{end}}=40$ months. The curve reaches its minimum at $\tau \approx 9$ months ($A(t) \approx -0.49$), then increases and plateaus near $A(t) \approx -0.39$ for $t > 30$ months. This non-monotonic $A(t)$ predicts that log-rank test power on data generated from similar survival functions would peak around 9 months and then stabilize. Extending follow-up beyond 24 months would yield little additional power because $A(t)$ has nearly reached its plateau by that time. For planning a future study with similar characteristics, this analysis suggests that investigators should consider alternative test statistics or, if using the log-rank test, recruiting additional patients to provide more power rather than following a fixed number of patients for longer.

\section{Discussion}

To our knowledge, this study provides the first broad evaluation of two-sample tests for survival in data with L-TS, with explicit attention to how the available follow-up time affects performance. We also provide a numerical tool for predicting the shape of the power curve over follow-up time for the conventional log-rank test in studies with L-TS.  We compared seven procedures, namely four log-rank type tests, two adaptive tests, and a correctly specified cure-model likelihood-ratio test, under hypotheses that target differences in both the uncured distribution and the cured fraction, in the uncured distribution only, or in the cured fraction.

For most candidates, type I error control was maintained at the nominal level. Across scenarios, the four log-rank tests and two adaptive tests exhibit largely similar power patterns as a function of follow-up time. When one or both groups lack L-TS, these hazard-based tests are typically the most powerful, with the ordering depending on sample size and effect size. When both groups contain L-TS and there is an effect on the uncured survival, the power of these hazard-based tests is non-monotonic as follow-up increases, whereas the correctly specified cure-model likelihood-ratio test gains power monotonically and achieves the highest power at the longest follow-up. 

Similarity in the behavior of the hazard-based candidates is explained by the fact that all six test statistics are functions of the hazard difference $h_1(t) - h_0(t)$ between treatment arms, either directly through the log-rank score \eqref{eq:U} or through reweighted versions of the same integrand. In data with L-TS, both $h_0(t)$ and $h_1(t)$ converge to zero as $t$ increases (equation~\ref{eq:mixtureh}), so the hazard difference (the signal these tests detect) diminishes at longer follow-up regardless of the weighting scheme. This shared dependence on a vanishing hazard signal produces the common non-monotonic power trajectory observed in our simulations when both groups contain L-TS. In contrast, the mixture cure model likelihood-ratio test jointly estimates the cure fractions $\pi_g$ and the uncured survival parameters, so it continues to accumulate information about group differences. However, we observe lower empirical power with earlier follow-up, likely due to insufficient follow-up for cure modeling.

We derived a numerical approach to predict the shape of the power curve for log-rank tests and confirmed its predictions with empirical power. This approach can accommodate alternative weights and arbitrary hazard functions (additional examples for these are given in the Supplementary Material). However, our empirical link between power and the weighted average \(A(\tau)\) in \eqref{eq:Adef} with weight \(w(\tau)\) in \eqref{eq:wdef}, and the log-rank mean in \eqref{eq:LRmean}, relies on approximations. First, when the sample size is small, the large sample replacement \(Y_g(t)\approx n_g S_g(t)G(t)\) may be limited, risk-set randomness is large, and the statistic \(U(\tau)\) can deviate from its expectation in ways not captured by \(A(\tau)\). Second, we ignored the variance term \(V(\tau)\) to focus on patterns of the predicted shape of the power curve. While ignoring the variance should not typically affect the predicted shape, it affects the scale. If \(V(\tau)\) changes quickly with \(\tau\), the power curve may differ from the pattern of \(A(\tau)\). Despite these limitations and assumptions, we still see that empirical power is higher as $A(\tau)$ is farther from 0 in our studied scenarios, as predicted. Therefore, we believe that this approach can produce numerical results to initially inform method choices for a study, which can then be refined with study-specific considerations.

Several limitations of our study warrant consideration for future research. We did not evaluate the restricted mean survival time (RMST) or differences in survival probabilities at fixed time points. Results from \citet{Klinglmueller2025} did not appear to suggest that these methods would outperform the candidates considered here, and these alternatives require choosing a clinically meaningful time horizon, which is context-dependent and outside the scope of this single paper. Future work could study principled ways to select or calibrate the time horizon in settings with L-TS.

 We quantified follow-up based on the quantiles of the reference group uncured distribution. This facilitates identifying generalizable patterns because follow-up is distribution-agnostic, but study-specific calendar times corresponding to the quantiles will necessarily be distribution-specific. Fortunately, this can be straightforwardly replicated for study-specific planning, as reference group data are typically already used in the study planning process, so following the methods of this manuscript can support developing the study-specific patterns for studies with L-TS.

Throughout the power scenarios, we encoded treatment effects so that the treatment survival curve uniformly dominates the control survival curve, and survival functions do not cross. When survival functions cross, there is no longer an unambiguous superior treatment arm, so hypothesis testing requires context-dependent considerations.

We assumed equal sample sizes between treatment groups and did not examine covariate adjustment. Unequal allocation, stratification by prognostic factors, and covariate-adjusted test statistics can alter efficiency and change the relative ordering of methods in terms of power. A systematic assessment of these design and analysis choices is important for study-specific considerations but is outside the scope of this paper.

We considered cure fractions up to 60\% for this research. From exploratory investigations, when the cure fraction in the control arm is 80\% or greater, all tests considered here exhibit poor power. This setting requires further study because so few subjects experience the event that there is little information to detect a treatment effect.

We focused on mixture-cure formulations with a broad class of distributions, which are considered in the design of certain oncology trials and did not examine non-mixture cure models or more generalized distributions. Under this setup, simulations across these classes of distributions showed that correctly specified cure-model likelihood-ratio tests often had lower power than hazard-based tests with shorter follow-up or lack of L-TS, whereas hazard-based tests were outperformed with longer follow-up. 

Using the BEACON-Immuno trial (NCT02308527)  study, we reviewed how to contextualize our results for planning a future study for relapsed high-risk neuroblastoma patients. We observed that extending follow-up beyond 24 months for a fixed sample size may only yield limited increase in power when using the conventional log-rank test statistic. For presentation purposes, calculations assumed the fitted model is correct, which may be a strong assumption for the moderate sample size available. Actual study planning would require additional sensitivity analyses, review of alternative methodology, and clinical considerations.

This work has an important impact for those currently designing studies in situations with L-TS, and study statisticians can use this research to interpret planned sample sizes jointly with follow-up times to ensure study power can be optimized. Results show that, to our knowledge, no method performs uniformly well across all follow-up scenarios in settings with L-TS, but our results can support choosing suitable methodology for studies with L-TS based on study-specific considerations. Future work should focus on developing approaches that can maintain high power across the spectrum of L-TS scenarios under arbitrary follow-up.

\bibliographystyle{agsm}
\bibliography{references}

@article{Larkin2019,
  author    = {James Larkin and Vanna Chiarion-Sileni and Rene Gonzalez and Jean Jacques Grob and C. Lance Cowey and Christopher D. Lao and Dirk Schadendorf and Reinhard Dummer and Michael Smylie and Piotr Rutkowski and Grant R. Ferrucci and others},
  title     = {Five-Year Survival with Combined Nivolumab and Ipilimumab in Advanced Melanoma},
  journal   = {New England Journal of Medicine},
  year      = {2019},
  volume    = {381},
  number    = {16},
  pages     = {1535--1546}
 
}

@article{Wolchok2022,
  author    = {Jedd D. Wolchok and Vanna Chiarion-Sileni and Rene Gonzalez and Jean Jacques Grob and Piotr Rutkowski and C. Lance Cowey and Christopher D. Lao and Dirk Schadendorf and Grant R. Ferrucci and Michael Smylie and Reinhard Dummer and others},
  title     = {Long-Term Outcomes With Nivolumab Plus Ipilimumab or Nivolumab Alone Versus Ipilimumab in Patients With Advanced Melanoma},
  journal   = {Journal of Clinical Oncology},
  year      = {2022},
  volume    = {40},
  number    = {2},
  pages     = {127--137}
}

@article{Tai2005,
  author    = {Patricia Tai and Edward Yu and G{\'a}bor Cserni and Georges Vlastos and Melanie Royce and Ian Kunkler and Vincent Vinh-Hung},
  title     = {Minimum Follow-Up Time Required for the Estimation of Statistical Cure of Cancer Patients: Verification Using Data from 42 Cancer Sites in the {SEER} Database},
  journal   = {BMC Cancer},
  year      = {2005},
  volume    = {5},
  number     = {48}
}

@article{Sposto2002,
  author    = {Richard Sposto},
  title     = {Cure Model Analysis in Cancer: An Application to Data from the {Children's Cancer Group}},
  journal   = {Statistics in Medicine},
  year      = {2002},
  volume    = {21},
  number    = {2},
  pages     = {293--312}
}

@article{Berkson1952,
  author    = {Joseph Berkson and Robert P. Gage},
  title     = {Survival Curve for Cancer Patients Following Treatment},
  journal   = {Journal of the American Statistical Association},
  year      = {1952},
  volume    = {47},
  number    = {259},
  pages     = {501--515}
}

@article{Kuk1992,
  author    = {Anthony Y. C. Kuk and Chi-Hse Chen},
  title     = {A Mixture Model Combining Logistic Regression with Proportional Hazards Regression},
  journal   = {Biometrika},
  year      = {1992},
  volume    = {79},
  number    = {3},
  pages     = {531--541},
  doi       = {10.1093/biomet/79.3.531}
}

@article{Sy2000,
  author    = {Judy P. Sy and Jeremy M. G. Taylor},
  title     = {Estimation in a {Cox} Proportional Hazards Cure Model},
  journal   = {Biometrics},
  year      = {2000},
  volume    = {56},
  number    = {1},
  pages     = {227--236},
  doi       = {10.1111/j.0006-341X.2000.00227.x},
  pmid      = {10783800}
}

@article{Peng2000,
  author    = {Yingwei Peng and Keith B. G. Dear},
  title     = {A Nonparametric Mixture Model for Cure Rate Estimation},
  journal   = {Biometrics},
  year      = {2000},
  volume    = {56},
  number    = {1},
  pages     = {237--243},
  doi       = {10.1111/j.0006-341X.2000.00237.x},
  pmid      = {10783801}
}

@article{Chen1999,
  author    = {Ming-Hui Chen and Joseph G. Ibrahim and Debajyoti Sinha},
  title     = {A New {Bayesian} Model for Survival Data with a Surviving Fraction},
  journal   = {Journal of the American Statistical Association},
  year      = {1999},
  volume    = {94},
  number    = {447},
  pages     = {909--919},
  doi       = {10.1080/01621459.1999.10474196}
}

@article{Tsodikov1998,
  author    = {Alexander Tsodikov},
  title     = {A Proportional Hazards Model Taking Account of Long-Term Survivors},
  journal   = {Biometrics},
  year      = {1998},
  volume    = {54},
  number    = {4},
  pages     = {1508--1516},
  doi       = {10.2307/2533675},
  pmid      = {9883549}
}

@article{Tsodikov2003,
  author    = {Alexander D. Tsodikov and Joseph G. Ibrahim and Alex Y. Yakovlev},
  title     = {Estimating Cure Rates from Survival Data: An Alternative to Two-Component Mixture Models},
  journal   = {Journal of the American Statistical Association},
  year      = {2003},
  volume    = {98},
  number    = {464},
  pages     = {1063--1078},
  doi       = {10.1198/01622145030000001007}
}

@article{Patilea2020,
  author    = {Valentin Patilea and Ingrid {Van Keilegom}},
  title     = {A General Approach for Cure Models in Survival Analysis},
  journal   = {The Annals of Statistics},
  year      = {2020},
  volume    = {48},
  number    = {4},
  pages     = {2323--2346},
  doi       = {10.1214/19-AOS1889}
}

@article{Kim2012,
  author    = {Haesook Teresa Kim and Robert Gray},
  title     = {Three-Component Cure Rate Model for Nonproportional Hazards Alternative in the Design of Randomized Clinical Trials},
  journal   = {Clinical Trials},
  year      = {2012},
  volume    = {9},
  number    = {2},
  pages     = {155--163},
  doi       = {10.1177/1740774512436614},
  pmid      = {22353928}
}

@article{Wei2020,
  author    = {Jing Wei and Jianrong Wu},
  title     = {Cancer Immunotherapy Trial Design with Cure Rate and Delayed Treatment Effect},
  journal   = {Statistics in Medicine},
  year      = {2020},
  volume    = {39},
  number    = {6},
  pages     = {698--708},
  doi       = {10.1002/sim.8440},
  pmid      = {31773770}
}

@article{Wu2022,
  author    = {Jianrong Wu and Jing Wei},
  title     = {Cancer Immunotherapy Trial Design with Random Delayed Treatment Effect and Cure Rate},
  journal   = {Statistics in Medicine},
  year      = {2022},
  volume    = {41},
  number    = {4},
  pages     = {786--797},
  doi       = {10.1002/sim.9258},
  pmid      = {34779534}
}

@article{Royston2013,
  author    = {Patrick Royston and Mahesh K. B. Parmar},
  title     = {Restricted Mean Survival Time: An Alternative to the Hazard Ratio for the Design and Analysis of Randomized Trials with a Time-to-Event Outcome},
  journal   = {BMC Medical Research Methodology},
  year      = {2013},
  volume    = {13},
  pages     = {152},
  doi       = {10.1186/1471-2288-13-152},
  pmid      = {24314264}
}

@article{Uno2014,
  author    = {Hajime Uno and Brian Claggett and Lu Tian and Eisuke Inoue and Paul Gallo and Toshio Miyata and Deborah Schrag and Masahiro Takeuchi and Yoshiaki Uyama and Lihui Zhao and Hicham Skali and Scott Solomon and Susanna Jacobus and Michael Hughes and Milton Packer and Lee-Jen Wei},
  title     = {Moving Beyond the Hazard Ratio in Quantifying the Between-Group Difference in Survival Analysis},
  journal   = {Journal of Clinical Oncology},
  year      = {2014},
  volume    = {32},
  number    = {22},
  pages     = {2380--2385},
  doi       = {10.1200/JCO.2014.55.2208},
  pmid      = {24982461}
}

@book{Fleming1991,
  author    = {Thomas R. Fleming and David P. Harrington},
  title     = {Counting Processes and Survival Analysis},
  publisher = {John Wiley \& Sons},
  address   = {New York},
  year      = {1991},
  series    = {Wiley Series in Probability and Statistics},
  isbn      = {0-471-52218-X}
}

@article{Royston2016,
  author    = {Patrick Royston and Mahesh K. B. Parmar},
  title     = {Augmenting the Logrank Test in the Design of Clinical Trials in Which Non-Proportional Hazards of the Treatment Effect May Be Anticipated},
  journal   = {BMC Medical Research Methodology},
  year      = {2016},
  volume    = {16},
  pages     = {16},
  doi       = {10.1186/s12874-016-0110-x},
  pmid      = {26869168}
}

@article{Mukhopadhyay2022,
  author    = {Pralay Mukhopadhyay and Jiabu Ye and Keaven M. Anderson and Satrajit Roychoudhury and Eric H. Rubin and Susan Halabi and Richard J. Chappell},
  title     = {Log-Rank Test vs {MaxCombo} and Difference in Restricted Mean Survival Time Tests for Comparing Survival Under Nonproportional Hazards in Immuno-oncology Trials: A Systematic Review and Meta-analysis},
  journal   = {JAMA Oncology},
  year      = {2022},
  volume    = {8},
  number    = {9},
  pages     = {1294--1300},
  doi       = {10.1001/jamaoncol.2022.2259},
  pmid      = {35862037}
}

@article{Klinglmueller2025,
  author    = {Florian Klinglm{\"u}ller and Tobias Fellinger and Franz K{\"o}nig and Tim Friede and Andrew C. Hooker and Harald Heinzl and Martina Mittlb{\"o}ck and Jonas Brugger and Maximilian Bardo and Cynthia Huber and Norbert Benda and Martin Posch and Robin Ristl},
  title     = {A Comparison of Statistical Methods for Time-to-Event Analyses in Randomized Controlled Trials Under Non-Proportional Hazards},
  journal   = {Statistics in Medicine},
  year      = {2025},
  volume    = {44},
  number    = {5},
  pages     = {e70019},
  doi       = {10.1002/sim.70019},
  pmid      = {39973243}
}

@article{Betensky2015,
  author    = {Rebecca A. Betensky},
  title     = {Measures of Follow-Up in Time-to-Event Studies: Why Provide Them and What Should They Be?},
  journal   = {Clinical Trials},
  year      = {2015},
  volume    = {12},
  number    = {4},
  pages     = {403--408},
  doi       = {10.1177/1740774515586176}
}

@article{Schemper1996,
  author    = {Michael Schemper and Trevor L. Smith},
  title     = {A Note on Quantifying Follow-Up in Studies of Failure Time},
  journal   = {Controlled Clinical Trials},
  year      = {1996},
  volume    = {17},
  number    = {4},
  pages     = {343--346},
  doi       = {10.1016/0197-2456(96)00075-X},
  pmid      = {8889347}
}

@article{Clark2003,
  author    = {Taane G. Clark and Michael J. Bradburn and Sharon B. Love and Douglas G. Altman},
  title     = {Survival Analysis Part {I}: Basic Concepts and First Analyses},
  journal   = {British Journal of Cancer},
  year      = {2003},
  volume    = {89},
  number    = {2},
  pages     = {232--238},
  doi       = {10.1038/sj.bjc.6601118}
}

@article{Gebski2018,
  author    = {Val Gebski and Val{\'e}rie Gar{\`e}s and Emma Gibbs and Karen Byth},
  title     = {Data Maturity and Follow-Up in Time-to-Event Analyses},
  journal   = {International Journal of Epidemiology},
  year      = {2018},
  volume    = {47},
  number    = {3},
  pages     = {850--859},
  doi       = {10.1093/ije/dyy013},
  pmid      = {29444326}
}

@article{Othus2020,
  author    = {Megan Othus and Aasthaa Bansal and Harry Erba and Scott Ramsey},
  title     = {Bias in Mean Survival from Fitting Cure Models with Limited Follow-Up},
  journal   = {Value in Health},
  year      = {2020},
  volume    = {23},
  number    = {8},
  pages     = {1034--1039},
  doi       = {10.1016/j.jval.2020.02.015},
  pmid      = {32828215}
}

@article{Maller1992,
  author    = {Ross A. Maller and Xian Zhou},
  title     = {Estimating the Proportion of Immunes in a Censored Sample},
  journal   = {Biometrika},
  year      = {1992},
  volume    = {79},
  number    = {4},
  pages     = {731--739},
  doi       = {10.1093/biomet/79.4.731}
}

@book{Maller1996,
  author    = {Ross A. Maller and Xian Zhou},
  title     = {Survival Analysis with Long-Term Survivors},
  publisher = {John Wiley \& Sons},
  address   = {New York},
  year      = {1996},
  isbn      = {0-471-96201-6}
}

@article{Selukar2023,
  author    = {Subodh Selukar and Megan Othus},
  title     = {{RECeUS}: Ratio Estimation of Censored Uncured Subjects, a Different Approach for Assessing Cure Model Appropriateness in Studies with Long-Term Survivors},
  journal   = {Statistics in Medicine},
  year      = {2023},
  volume    = {42},
  number    = {3},
  pages     = {209--227},
  doi       = {10.1002/sim.9610},
  pmid      = {36433635}
}

@article{Xiong2017,
  author    = {Xiaoping Xiong and Jianrong Wu},
  title     = {A Novel Sample Size Formula for the Weighted Log-Rank Test Under the Proportional Hazards Cure Model},
  journal   = {Pharmaceutical Statistics},
  year      = {2017},
  volume    = {16},
  number    = {1},
  pages     = {87--94},
  doi       = {10.1002/pst.1790},
  pmid      = {27860138}
}

@book{Peng2021,
  author    = {Yingwei Peng and Binbing Yu},
  title     = {Cure Models: Methods, Applications, and Implementation},
  publisher = {Chapman and Hall/CRC},
  address   = {Boca Raton, FL},
  year      = {2021},
  series    = {Chapman \& Hall/CRC Biostatistics Series},
  isbn      = {978-0-367-14557-6}
}

@article{Schoenfeld1981,
  author    = {David A. Schoenfeld},
  title     = {The Asymptotic Properties of Nonparametric Tests for Comparing Survival Distributions},
  journal   = {Biometrika},
  year      = {1981},
  volume    = {68},
  number    = {1},
  pages     = {316--319},
  doi       = {10.1093/biomet/68.1.316}
}

@article{Boag1949,
  author    = {John W. Boag},
  title     = {Maximum Likelihood Estimates of the Proportion of Patients Cured by Cancer Therapy},
  journal   = {Journal of the Royal Statistical Society: Series B (Methodological)},
  year      = {1949},
  volume    = {11},
  number    = {1},
  pages     = {15--53},
  doi       = {10.1111/j.2517-6161.1949.tb00020.x}
}

@article{Amico2018,
  author    = {Ma{\"i}lis Amico and Ingrid {Van Keilegom}},
  title     = {Cure Models in Survival Analysis},
  journal   = {Annual Review of Statistics and Its Application},
  year      = {2018},
  volume    = {5},
  pages     = {311--342},
  doi       = {10.1146/annurev-statistics-031017-100101}
}

@article{Kouadio2025,
  author  = {Kouadio, Cheryl and Selukar, Subodh and Othus, Megan and Chevret, Sylvie},
  title   = {Detecting the Cure Model Appropriateness in Randomized Clinical Trials With Long-Term Survivors},
  journal = {JCO Clinical Cancer Informatics},
  year    = {2025},
  volume  = {9},
  pages   = {e2500084},
  doi     = {10.1200/CCI-25-00084}
}

@article{Gray2026,
  author  = {Gray, Juliet C. and Weston, Rebekah and Owens, Cormac and Canete, Adela and Gambart, Marion and De Wilde, Bram and Nysom, Karsten and van Eijkelenburg, Natasha and Ladenstein, Ruth and Castellano, Aurora and Gerber, Nicolas U. and Marshall, Lynley V. and Barone, Giuseppe and Rubio-San-Simon, Alba and Ng, Antony and Vaidya, Sucheta and Gallego, Soledad and Makin, Guy and Burke, G. A. Amos and McCarthy, Anthony and Murphy, Dermot and Zwaan, C. Michel and L{\'o}pez-Almaraz, Ricardo and Jannier, Sarah and Thebaud, Estelle and Corradini, Nadege and Yeomanson, Dan and Howell, Lisa and Tweddle, Deborah A. and Elliott, Martin and Hobin, Dave and Valteau-Couanet, Dominique and Schleiermacher, Gudrun and Chastagner, Pascal and Defachelles, Anne Sophie and Brichard, Benedicte and George, Sally and Chesler, Louis and Laidler, Jennifer and Firth, Charlotte and Holt, Grace and Moroz, Veronica and Pearson, Andrew D. J. and Gates, Simon and Wheatley, Keith and Kearns, Pam and Moreno, Lucas},
  title   = {Dinutuximab Beta Added to Temozolomide-Based Chemotherapy for Children With Relapsed and Refractory Neuroblastoma: Results of the {ITCC-SIOPEN} {BEACON} Immuno Phase {II} Trial},
  journal = {Journal of Clinical Oncology},
  year    = {2026},
  volume  = {44},
  number  = {3},
  pages   = {176--187},
  doi     = {10.1200/JCO-25-01868}
}

@article{YangPrentice2010,
  author    = {Yang, S. and Prentice, R.},
  title     = {Improved Logrank-Type Tests for Survival Data Using Adaptive Weights},
  journal   = {Biometrics},
  year      = {2010},
  volume    = {66},
  pages     = {30--38},
  doi       = {10.1111/j.1541-0420.2009.01243.x}
}

@article{QiuSheng2008,
  author    = {Qiu, P. and Sheng, J.},
  title     = {A two-stage procedure for comparing hazard rate functions},
  journal   = {Journal of the Royal Statistical Society: Series B (Statistical Methodology)},
  year      = {2008},
  volume    = {70},
  pages     = {191--208},
  doi       = {10.1111/j.1467-9868.2007.00622.x}
}

\begin{table}[h]
\centering
\begin{threeparttable}
\caption{Comparison of Survival Hypothesis Tests}
\label{tab:survival_comparison}
\small
\begin{tabular}{lp{4.5cm}ccc}
\toprule
\textbf{Test} & \textbf{Test Statistic / Procedure} & \textbf{Hazard Test} & \textbf{NPH Design} & \textbf{Parametric} \\
\midrule
Conventional log-rank & Unweighted ($w=1$) & Y & N & N \\ \addlinespace
Mixture Cure & Likelihood ratio test & N & Y & Y \tnote{a} \\  \addlinespace
Early WLR & FH $G^{1,0}$; higher weights for early events. & Y & Y & N \\ \addlinespace
Late WLR & FH $G^{0,1}$; higher weights for late events. & Y & Y & N \\ \addlinespace
Optimal LR & $w \propto \log(\text{HR})$; where $\rho \approx -1, \vartheta \approx 0$. & Y & Y & N \\ \addlinespace
Adaptive YP & Allows a flexible HR change over time by distinguishing $\theta_E$ (short) and $\theta_L$ (long) HRs. & Y & Y & Semiparametric \\ \addlinespace
Two-Stage (TS) & Stage 1: Log-rank; Stage 2: FH weights + Bootstrap. & Y & Y & N \\
\bottomrule
\end{tabular}
\begin{tablenotes}
\footnotesize
\item[a] \textit{In our simulation, we used a correctly-specified parametric mixture cure model; however, semiparametric and nonparametric mixture cure model formulas are also available in the literature.}
\end{tablenotes}
\end{threeparttable}
\end{table}

\begin{table}[ht]
\centering
\caption{Summary of Simulation Study Design Parameters}
\label{tab:sim_parameters}
\small
\begin{tabular}{lll}
\toprule
\textbf{Category} & \textbf{Factor} & \textbf{Values / Settings} \\
\midrule
\textbf{Distributions} & Uncured Survival ($S_{u,0}$) & Weibull $W(k=2, \lambda=1)$ \\
&  & Log-logistic $LL(\beta=2, \alpha=1)$ \\
&  & Gamma $Gamma(\alpha=2, \lambda=1)$ \\
& Cure Link Function & Logit: $\ln(\pi / (1-\pi))$ \\
\addlinespace
\textbf{Effect Sizes} & Odds Ratio (OR) & $1.0, 1.1, 1.25, 1.5$ \\
& Uncured HR ($HR_u$) & $1.0, 0.9, 0.75, 0.5$ for Weibull\\
& Corresponding TR & $1.0, 1.05, 1.15, 1.41$ for LL and Gamma\\
\addlinespace
\textbf{Cure Fractions} & Control group $\pi_0$ & $0, 0.1, 0.2, 0.5, 0.8, 0.95$ \\
\addlinespace
\textbf{Study Design} & Sample size ($n$) & $25, 50, 100, 500$ per arm \\
& Follow-up ($\tau$) & Quantiles: $0.75, 0.9, 0.95, 0.99, 0.999$ \\
& Accrual Duration & Fixed at up to half of 75th quantile of $S_{u,0}$ \\
& Time Unit & Rounded to nearest quarter-year \\
\addlinespace
\textbf{Computational} & Iterations & 10,000 per scenario \\
& Platform & High Performance Computing (HPC) \\
\bottomrule
\end{tabular}
\end{table}

\end{document}